\documentclass[conference]{IEEEtran}
\usepackage[ruled,vlined]{algorithm2e}

\ifCLASSINFOpdf

\else

\fi

\usepackage{amsfonts,epsfig,cite,array,multirow,graphicx,amsmath,ltablex,tabularx,setspace,amssymb,multirow}
\usepackage{schemabloc,tikz,amsthm}
\usepackage{subfig}
\usepackage[numbers,sort&compress]{natbib}
\usetikzlibrary{circuits, arrows}
\usepackage{verbatim}
\usepackage{graphicx}
\usepackage{epstopdf}
\usepackage{mathtools}
\usepackage{xcolor}
\usepackage{marvosym}

\usepackage{lipsum}
\usepackage{dblfloatfix}
\usepackage{color}
\usepackage{etoolbox}
\usepackage{pstricks}
\usepackage{epsfig}
\theoremstyle{plain}

\newtheorem{exm}{Example}
\renewcommand\qedsymbol{$\blacksquare$}
\usepackage{tikz}
\usetikzlibrary{shapes}
\usetikzlibrary{plotmarks}
\usepackage{pgfplots}
\usetikzlibrary{spy}
\usepackage{collectbox}

\makeatletter

\begin{document}
	\tikzset{new spy style/.style={spy scope={%
				magnification=2.8,
				size=1.25cm,
				connect spies,
				every spy on node/.style={
					rectangle,
					draw,
				},
				every spy in node/.style={
					draw,
					rectangle,
				}
			}
		}
	}
	\SetKwRepeat{Do}{do}{while}
	\newcolumntype{L}[1]{>{\raggedright\let\newline\\\arraybackslash\hspace{0pt}}m{#1}}
	\newcolumntype{C}[1]{>{\centering\let\newline\\\arraybackslash\hspace{0pt}}m{#1}}
	\newcolumntype{R}[1]{>{\raggedleft\let\newline\\\arraybackslash\hspace{0pt}}m{#1}}
	\pgfdeclarelayer{background}
	\pgfdeclarelayer{foreground}
	\pgfsetlayers{background,main,foreground}
	\newcommand{\oeq}{\mathrel{\text{\sqbox{$=$}}}}
	\setlength{\textfloatsep}{0.1cm}
	\setlength{\floatsep}{0.1cm}
	\tikzstyle{int}=[draw, fill=white!20, minimum size=2em]
	\tikzstyle{init} = [pin edge={to-,thin,black}]

	\title{Sparse Code  Multiple Access (SCMA) Technique}
	
	\author{
		\IEEEauthorblockN{ Sanjeev Sharma and Kuntal Deka\\IIT (BHU) Varanasi,  India. sanjeevs.ece@iitbhu.ac.in\\		IIT Guwahati,  India. kuntal@iitg.ac.in }}
	\maketitle
	

{\abstract{Next-generation wireless networks require higher spectral efficiency and lower latency to meet the demands of various upcoming  applications. Recently, non-orthogonal multiple access (NOMA) schemes are introduced in the literature for 5G and beyond. Various forms of NOMA are considered like power domain, code domain, pattern division multiple access, etc. to enhance the spectral efficiency of  wireless networks.
	In this chapter, we introduce the code domain-based sparse code multiple access (SCMA) NOMA scheme to enhance the spectral efficiency of a wireless  network. The design and detection of an SCMA system are  analyzed in this  chapter. Also, the method for codebooks design and its impact on  system performance are  highlighted.  A hybrid multiple access scheme is also introduced using both code-domain and power-domain NOMA. Furthermore, simulation results are included to show  the impact of various SCMA system parameters.}}

\section{Introduction of NOMA}
In this chapter, we focus on the sparse code multiple access (SCMA) and hybrid multiple access (HMA) schemes for  next-generation wireless systems.
From 1G to 4G wireless systems are based on the orthogonal multiple access (OMA) techniques such as time division multiple access (TDMA) and frequency division multiple access (FDMA), where resources are allocated to  each user exclusively. 
OMA-based system has lower  spectral efficiency  when some bandwidth resources are allocated to users with poor channel state information (CSI). Further, OMA  techniques  may not be suitable for a  wireless network which requires very high spectral efficiency, very low latency, and massive device connectivity. Therefore, OMA based system may not be able to support the  connectivity of  billions devices in wireless network for various futuristic applications.  Hence,  researchers have started to focus on NOMA-based systems to achieve higher spectral efficiency and lower latency   for next-generation wireless networks.  In NOMA,  one or more resources are allocated to more than one user simultaneously to enhance the spectral efficiency of network. Further, power domain (PD) and code domain NOMA-based  systems are mainly explored in the literature \cite{Dai, sharma, ping}. Recently, NOMA-based systems are also analyzed by considering the multiple-input multiple-output (MIMO) techniques \cite{CHENG, Zhipeng, Weijie}.

SCMA is a code domain (CD) NOMA technique, in which a user occupies more than one orthogonal resource\footnote{orthogonal time slots or frequency bands or codes are referred to as  orthogonal resource in the system. } for communication. In SCMA, multi-dimensional codebooks are used  due to which shaping gain can be enhanced. Further, SCMA can be thought of as an extended version of low density spreading (LDS) multiple access technique. 
Recently, SCMA has brought a significant   interest from the researchers and the scientists  for 5G and beyond wireless networks.

\section{SCMA System Model}
\begin{figure}[htb!]
	\centering
	\includegraphics[]{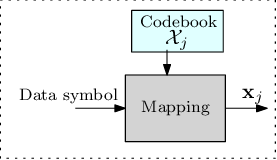}
	\caption {Data symbol to codeword mapping. }	 	
	\label{mapping}
\end{figure}

Suppose the number of orthogonal resource elements and   users in a system  are $K$ and $J$, where $J>K$. Therefore, $J$ users communicate using $K$ resources and this system is often referred to as $J\times K$ SCMA system.   In SCMA, each user's information is directly mapped to a multidimensional complex vector for transmission.  Let each user has a codebook $\mathcal{X}_{j}, j=1,2,...,J$, which contains $M$ complex columns of dimension $K$. Therefore, each user codebook $\mathcal{X}_{j}$  has a  size is $K \times M $ with complex elements and can be written as $\mathcal{X}_{j}=[\begin{matrix}
	\mathbf{x}_{j1}, &\mathbf{x}_{j2},& \cdots, &\mathbf{x}_{jM}
\end{matrix} ]\in \mathbb{C}^{K\times M}$. For example, $M=4$, each data symbol has two bits, i.e., $\log_{2}(M)$ bits. Each column of codebook $\mathcal{X}_{j}$ corresponds to a data symbol. Hence, first, second, third, and fourth columns are selected corresponding data symbols $[00]$, $[01]$, $[10]$, and $[11]$, respectively,  for transmission. The encoding operation of an SCMA system  is illustrated  in Fig. \ref{mapping}.

\begin{figure*}[htb!]
	\centering
	\includegraphics[width=\linewidth]{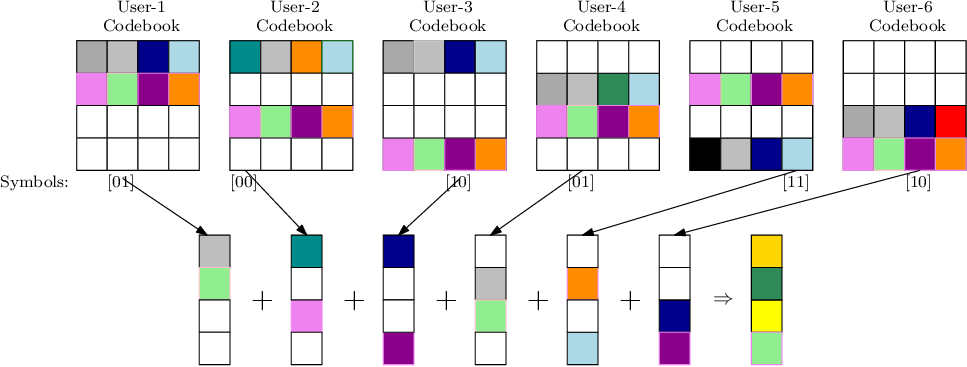}
	\caption {SCMA codebook model  for $J=6$ and $K=4$. }	 	
	\label{fig2}
\end{figure*}

Consider the example of  $6 \times 4$ SCMA system.   The  6  codebooks and their superposition are depicted  in Fig. \ref{fig2}. Each codebook has four columns corresponding  to the four data  symbols. Further, only two same rows are non-zeros in each codebook. Therefore, the  codebooks are sparse\footnote{Only a few elements in a signal are non-zeros as compared to the total number of elements}. The sparsity of the codebooks can help to achieve low complexity detection of users symbols. Each user selects a column from their codebooks, and all the codewords are summed, as shown in Fig. \ref{fig2}. The values of the non-zero elements and their locations in the codebooks can be selected in  such a way that  the  system performance is optimized. 
Next section, we briefly summarize the downlink and the  uplink  SCMA system model.

\subsubsection{Downlink SCMA system}
\begin{figure*}[htb!]
	\centering
	\includegraphics[width=\linewidth]{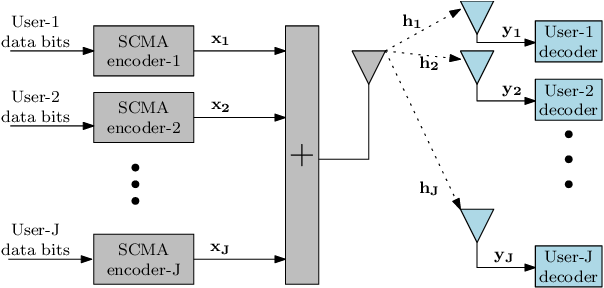}
	\caption {Downlink  SCMA system. }	 	
	\label{downlink}
\end{figure*}

In downlink SCMA, all users' information is  broadcast  from the  base station to 
and each user node receives the sum of all users' codewords.
The received signal ${\mathbf{y}}_i=\left[y_{i1}, \ldots,  y_{iK}\right]$ at the $i$th user can be expressed as \cite{deka}
\begin{equation}\label{int1}
	\mathbf{y}_i=\text{diag}\left({\bf{h}}_i\right)\sum_{j=1}^{J}  \sqrt{P_{j}} \mathbf{x}_{j}+\mathbf{n}_i,
\end{equation} 
where ${\bf{h}}_i=\left[h_{i1}, h_{i2}, \ldots, h_{iK}\right]$ is the channel impulse response  vector at the $i$th user and 
$P_{j}$ is the  power assigned to  the $j$th user.  $\mathbf{x}_{j}=[x_{j1}, x_{j2},...,x_{jK}] \in \mathbb{C}^{K}$ is the $j$th SCMA codeword   and  $\mathbf{n}_i$ is the   additive white Gaussian noise (AWGN) at the $i$th user and is Gaussian distributed, i.e., $\mathbf{n}_i\sim \mathcal{CN}(0,N_{0}\mathbf{I}_K)$. A downlink  $J \times K$ SCMA system is shown in Fig. \ref{downlink}. Message passing algorithm (MPA) is employed by  each user to decode the data symbol.

\subsubsection{Uplink SCMA system}
\begin{figure*}[htb!]
	\centering
	\includegraphics[width=\linewidth]{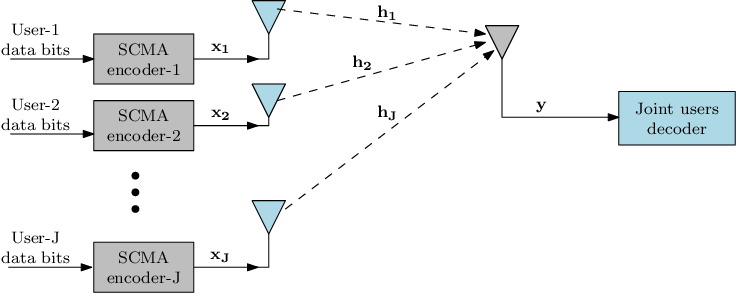}
	\caption{ Uplink  SCMA system. }	 	
	\label{uplink}
\end{figure*}

In uplink SCMA, each user access the respective channel to transmit information  to the base station, as shown in Fig. \ref{uplink}.
Consider a system  involving  $J$ users sharing   $K$ orthogonal  resource elements.
Each user has a codebook ${\cal{X}}_j$ which contains $M$, $K$-dimensional constellations: ${\cal{X}}_j= \left\{{\bf{x}}_{j1}, {\bf{x}}_{j2}, \ldots, {\bf{x}}_{jM}\right\}$.
In Rayleigh fading the received  signal $\textbf{y}$ is expressed as
\begin{equation}\label{ray1}
	\textbf{y}=\sum_{j=1}^{J} \sqrt{P_{j}} \operatorname{diag(\textbf{h}_{j})}\textbf{x}_{j}+\textbf{n},
\end{equation}
where $\textbf{h}_{j}=[h_{1j}, h_{2j},\ldots,h_{Kj}] \in \mathbb{C}^{K}$ is the channel impulse response  vector between the receiver and $j$th user. The envelope of each  $h_{jk}, k=1,2,...,K$ is Rayleigh distributed. $P_j$ is the power of the $j$th user. In uplink SCMA system users' data symbol can be  decoded jointly using the MPA at the base station.

\subsection{Design and Optimization of SCMA Codebooks}
The performance of an SCMA system  depends on the codebooks  and their assignment to the users. The methods to design optimum  codebooks in guaranteed fashion  do not exist  in literature.   Usually the codebooks   are designed using sub-optimal methods. For example in  \cite{zhang}, codebooks are designed by optimizing   the single dimensional complex code-words instead using the multidimensional  codewords. Similarly, in \cite{Lisu}, quadrature amplitude modulation and phase rotation based codebooks are designed using sub-optimal method.

The codebook design problem can be formulated by maximizing the sum rate of the users as  \cite{zhang}
\begin{equation}\label{sum1}
	R_{\mathrm s}\leq I(\mathbf{y};\mathcal{X}_{1}, \mathcal{X}_{2},..., \mathcal{X}_{J} ),
\end{equation}
where $\mathbf{y}$ is the $K$ dimensional received signal vector and $I(\cdot)$ is the  mutual information between the received signal $\mathbf{y}$ and multiuser codewords $\{\mathcal{X}_{j}\}_{j=1}{J}$. Therefore, the codebooks are optimized as
\begin{equation}\label{sum2}
	\{\mathcal{X}^{\mathrm{opt}}_{1}, \mathcal{X}^{\mathrm{opt}}_{2},..., \mathcal{X}^{\mathrm{opt}}_{J} \}=\arg \max_{\mathcal{X}_{j}} I(\mathbf{y};\mathcal{X}_{1}, \mathcal{X}_{2},..., \mathcal{X}_{J} ).
\end{equation}
Hence, the set of codewords $\{\mathcal{X}^{\mathrm{opt}}_{1}, \mathcal{X}^{\mathrm{opt}}_{2},..., \mathcal{X}^{\mathrm{opt}}_{J} \}$ represents the optimal value of codebooks for which the sum rate of users $R_{\mathrm s}$ has the maximum value.

\begin{exm}
	In this example, a codebook design method is illustrated for each user by considering $6\times 4$ SCMA system. Codebooks are designed by maximizing the mutual information between the
	received signal $\mathbf{y}$ and interfering users' intermingled data, and
	the shaping gain of the constellation points \cite{sharma}. Let the codewords of $d_{f}$ users  are summed at each resource node for transmission. From the summed codeword, each user information
	has to  be recovered. The constellation points must
	be designed in such a way that the sums are distinct.
	
	Let  $Y$ be  the received signal over one resource element and $S$ be  the sum of the codewords of  $d_f$  users.  The alphabet $\cal{S}$ for $S$ contains $M^{d_f}$ distinct sum values and is denoted by
	$$
	{\cal{S}}= \left\{s_1, s_2, \ldots, s_{M^{d_f}}\right\}
	$$
	where, $M$ is the modulation order and  $\left\{s_i\right\}_{i=1}^{M^{d_f}}$s is the set of  distinct sum values. Mutual information between $Y$ and $S$ is expressed \cite{sharma} as
	\begin{equation}
		\label{capacity}
		\begin{split}
			I_{m}\left(Y;S\right)& = \log |{\cal{S}}| -\frac{1}{|\cal{S}|} \sum_{j=1}^{{|\cal{S}}|} \frac{1}{\pi N_0}
			\int_{y \in \mathbb{C}}
			\exp \left(-\frac{\left\Vert y-s_j \right \Vert^2}{N_0}\right)\\
			& \times \log \left[\sum_{\substack{i=1 }}^{|\cal{S}|} \exp \left(\frac{\left\Vert y-s_j\right \Vert^2-\left\Vert y-s_i\right \Vert^2}{N_0}\right)\right]dy.
		\end{split}
	\end{equation}
	However, optimization of mutual information $I_{m}\left(Y;S\right)$ is difficult and one can choose the lower bound of $I_{m}\left(Y;S\right)$ as \cite{sharma}
	\begin{equation}
		\label{lower_bound}
		\begin{split}
			I_{m}^L=& \log |{\cal{S}}| -\\
			& \log \left[1+\frac{1}{|{\cal{S}}|} \sum_{j=1}^{|{\cal{S}}|}
			\sum_{\substack{i=1 \\ i \neq j}}^{|{\cal{S}}|}\exp \left(-\frac{1}{4N_0} ||s_j-s_i||^2\right)\right].
		\end{split}
	\end{equation}
	
	Further, the shaping gain of the constellation points is optimized to get an improved performance of the codebooks.  It is widely accepted that a high value of shaping
	gain is obtained by adopting a nearly-circular constellation
	boundary. Moreover, the irregularity in the constellation points
	tends to increase the diversity, which results in a higher value
	of shaping gain \cite{sharma}.
	The shaping gain of a region ${\cal{R}}$ is expressed as  \cite{Forney}
	\begin{equation}
		\gamma_s\left({\cal{R}}\right) =\frac{[V\left({\cal{R}}\right)]^{\frac{2}{n}}}{6 {\cal{E}}_{\text{av}}}
	\end{equation}
	where, $V\left({\cal{R}}\right)$ is the volume of the region ${\cal{R}}$,  $n$ is the dimension  and ${\cal{E}}_{\text{av}}$ is the average energy of the constellation points.
	
	The minimum Euclidean
	distance $d_{\mathrm{min}}$ of the constellation points plays an important role in the design of codebooks apart from the shaping gain and mutual information. In this example, we considered Euclidean
	distance as unity. Therefore, a constraint on the constellation points is imposed as
	$$
	\mathcal{E}_{\mathrm {av}}=\frac{1}{M}\sum_{m=1}^{M} \lVert \boldsymbol {x}_{m} \rVert^{2}=1.
	$$
	Therefore, one can jointly  optimize mutual information, shaping gain, and
	the minimum Euclidean distance to design codebooks. However, this joint optimization is difficult and a  sub-optimal approach can be employed by optimizing   each one of the
	objectives separately.
	
	Let $M=4$, $N=2$ and $d_{f}=3$.  One can select $4$-points using pulse amplitude modulation (PAM) constellation points which are  equally spaced in line or space. These constellation points correspond to  the first user and is denoted as $\mathcal{U}_{1}$.  $\mathcal{U}_{1}$ points are expressed as  $\mathcal{U}_{1}=\{-1, -0.333, 0.333, 1\}$ and  these points are scaled
	to get the unit average energy. Points of other two interfering users $\mathcal{U}_{2}$ and $\mathcal{U}_{3}$ are generated by  judiciously rotating the constellation points $\mathcal{U}_{1}$.  The angles of
	$\mathcal{U}_{2}$ and $\mathcal{U}_{3}$   are generated  by maximizing the
	mutual information and the shaping gain: 
	\begin{equation}\label{opt3}
		\begin{split}
			\{\theta^{\star}_{2}, \theta^{\star}_{3}\}&=\arg  \max_{\{\theta_{2},\theta_{3}\} \in [0  \ 2\pi)} I_{m}^L \\
		\end{split}
	\end{equation}
	where $I_{m}^L$ is given in (\ref{lower_bound}). $\{\theta^{\star}_{2}, \theta^{\star}_{3}\}$ are generated using (\ref{opt3}) and  then the  shaping gain is optimized by considering the irregularity  in the constellations $\mathcal{U}_{2}$ and $\mathcal{U}_{3}$.
	Additionally, during  optimization, the Euclidean distance among the constellation points is kept as fixed. The complete codebook from  constellation points, ${\cal{U}}_1,{\cal{ U}}_2$  and ${\cal{U}}_3$ are distributed over the resources as
	$$
	\mathcal{I}=
	\begin{bmatrix}
		\mathcal{U}_{1} & 0 & \mathcal{U}_{2} & 0 & \mathcal{U}_{3} & 0\\
		0 & \mathcal{U}_{2} & \mathcal{U}_{3} & 0 & 0 & \mathcal{U}_{1}\\
		\mathcal{U}_{2} & 0 & 0 & \mathcal{U}_{1} & 0 & \mathcal{U}_{3}\\
		0 & \mathcal{U}_{1} & 0 & \mathcal{U}_{3} & \mathcal{U}_{2} & 0\\
	\end{bmatrix},
	$$
	where $\mathcal{I}$ indicator matrix. The values of optimum angles  $\theta^{\star}_{2}=60^{\circ}, \theta^{\star}_{3}=120^{\circ}$ are obtained at $10$ dB.  The generated constellation points are given as
	\begin{equation}
		\begin{split}
			\mathcal{U}_{1}=[-1, -0.333, 0.333, 1]\\
			\mathcal{U}_{2}=[-0.1109-0.3i, 0.6+1i,    -0.6-1i,   0.1109+0.3i]\\
			\mathcal{U}_{3}=[0.3-0.3i, -0.6+1i,     0.6-1i,     -0.3+.3i].
		\end{split}
	\end{equation}
	The  codebooks of all users  in the  $6 \times 4$  SCMA system  is expressed in Table~\ref{table:codebook}.
	\begin{table}[ht]
		\centering
		\caption{Designed codebooks of all $6$ users}
		\label{table:codebook}
		\scalebox{0.7}{$
			\begin{gathered}
				{\cal{X}}_1=\left\{\begin{bmatrix}
					-1 \\
					0\\
					-0.1109-0.3i\\
					0
				\end{bmatrix}
				\begin{bmatrix}
					-0.333 \\
					0\\
					0.6+1i\\
					0
				\end{bmatrix}
				\begin{bmatrix}
					0.333 \\
					0\\
					-0.6-1i\\
					0
				\end{bmatrix}
				\begin{bmatrix}
					1 \\
					0\\
					0.1109+0.3i\\
					0
				\end{bmatrix}  \right\} \\
				{\cal{X}}_2=\left\{\begin{bmatrix}
					0\\
					-0.1109-0.3i\\
					0\\
					-1
				\end{bmatrix}
				\begin{bmatrix}
					0\\
					0.6+1i\\
					0\\
					-0.333
				\end{bmatrix}
				\begin{bmatrix}
					0\\
					-0.6-1i\\
					0\\
					0.333
				\end{bmatrix}
				\begin{bmatrix}
					0\\
					0.1109+0.3i\\
					0\\
					1
				\end{bmatrix}   \right\} \\
				{\cal{X}}_3=\left\{\begin{bmatrix}
					-0.6-1i\\
					0.3-0.3i\\
					0\\
					0
				\end{bmatrix}
				\begin{bmatrix}
					-0.1109-0.3i\\
					-0.6+1i \\
					0\\
					0
				\end{bmatrix}
				\begin{bmatrix}
					0.1109+0.3i\\
					0.6-1i\\
					0\\
					0
				\end{bmatrix}
				\begin{bmatrix}
					0.6+1i\\
					-0.3+.3i\\
					0\\
					0
				\end{bmatrix}   \right\} \\
				{\cal{X}}_4=\left\{\begin{bmatrix}
					0\\
					0\\
					-1\\
					0.3-0.3i
				\end{bmatrix}
				\begin{bmatrix}
					0\\
					0\\
					-0.333 \\
					-0.6+1i
				\end{bmatrix}
				\begin{bmatrix}
					0\\
					0\\
					0.333\\
					0.6-1i
				\end{bmatrix}
				\begin{bmatrix}
					0\\
					0\\
					1\\
					-0.3+.3i
				\end{bmatrix}   \right\} \\
				{\cal{X}}_5=\left\{\begin{bmatrix}
					0.3-0.3i \\
					0\\
					0\\
					-0.6-1i
				\end{bmatrix}
				\begin{bmatrix}
					-0.6+1i\\
					0\\
					0\\
					-0.1109-0.3i
				\end{bmatrix}
				\begin{bmatrix}
					0.6-1i\\
					0\\
					0\\
					0.1109+0.3i
				\end{bmatrix}
				\begin{bmatrix}
					-0.3+.3i \\
					0\\
					0\\
					0.6+1i
				\end{bmatrix}   \right\} \\
				{\cal{X}}_6=\left\{\begin{bmatrix}
					0\\
					-1\\
					0.3-0.3i\\
					0
				\end{bmatrix}
				\begin{bmatrix}
					0\\
					-0.333 \\
					-0.6+1i \\
					0
				\end{bmatrix}
				\begin{bmatrix}
					0\\
					0.333 \\
					0.6-1i \\
					0
				\end{bmatrix}
				\begin{bmatrix}
					0\\
					1\\
					-0.3+.3i \\
					0
				\end{bmatrix}   \right\}
			\end{gathered}
			$}
	\end{table}
	
\end{exm}

\subsection{Symbol detection in SCMA}
In this section, 
the detection of the users (known as multi-user detection (MUD)) is described.  Usually message passing algorithm (MPA) is used for MUD.  The MPA can be best explained graphically in terms of the factor graph. Therefore, first we  explain the factor graph. 

\vspace{-.6em}
\subsubsection{Factor Graph Representation} \label{ffg}
\begin{figure}[!h]
	\begin{center}
		\scalebox{1.2}{
			\begin{tikzpicture}[node distance=2.5cm,auto,>=latex']
				\draw [] (-0.5,0) node (xaxis) [] {UN};
				\draw [] (0,0) node (xaxis) [] { \large{$\bigcirc$}};
				\draw [] (0,0) node (xaxis) [] {$1$};	
				\draw [] (1,0) node (xaxis) [] {\large{$\bigcirc$}};\
				\draw [] (1,0) node (xaxis) [] {$2$};
				\draw [] (2,0) node (xaxis) [] {\large{$\bigcirc$}};
				\draw [] (2,0) node (xaxis) [] {$3$};	
				\draw [] (3,0) node (xaxis) [] {\large{$\bigcirc$}};
				\draw [] (3,0) node (xaxis) [] {$4$};
				\draw [] (4,0) node (xaxis) [] {\large{$\bigcirc$}};
				\draw [] (4,0) node (xaxis) [] {$5$};	
				\draw [] (5,0) node (xaxis) [] {\large{$\bigcirc$}};
				\draw [] (5,0) node (xaxis) [] {$6$};
				\draw [] (0.2,-1.25) node (xaxis) [] {RN};
				\draw  [] (0.7,-1.25) node (xaxis) [] {\Large{$\square$}};
				\draw  [] (0.7,-1.25) node (xaxis) [] {$1$};		
				\draw [] (1.9,-1.25) node (xaxis) [] {\Large{$\square$}};
				\draw [] (1.9,-1.25) node (xaxis) [] {$2$};
				\draw [] (3.1,-1.25) node (xaxis) [] {\Large{$\square$}};
				\draw [] (3.1,-1.25) node (xaxis) [] {$3$};
				\draw [] (4.4,-1.25) node (xaxis) [] {\Large{$\square$}};
				\draw [] (4.4,-1.25) node (xaxis) [] {$4$};

				\draw[-] (0,-0.17) coordinate (a_1) -- (0.7,-1.13) coordinate (a_2);
				\draw[-] (4,-0.17) coordinate (a_1) -- (0.7,-1.13) coordinate (a_2);
				\draw[-] (2,-0.17) coordinate (a_1) -- (0.7,-1.13) coordinate (a_2);
				
				\draw[-] (1,-0.17) coordinate (a_1) -- (1.9,-1.13) coordinate (a_2);
				\draw[-] (2,-0.17) coordinate (a_1) -- (1.9,-1.13) coordinate (a_2);
				\draw[-] (5,-0.17) coordinate (a_1) -- (1.9,-1.13) coordinate (a_2);
				
				\draw[-] (0,-0.17) coordinate (a_1) -- (3.1,-1.13) coordinate (a_2);
				\draw[-] (3,-0.17) coordinate (a_1) -- (3.1,-1.13) coordinate (a_2);
				\draw[-] (5,-0.17) coordinate (a_1) -- (3.1,-1.13) coordinate (a_2);
				
				\draw[-] (1,-0.17) coordinate (a_1) -- (4.4,-1.13) coordinate (a_2);
				\draw[-] (3,-0.17) coordinate (a_1) -- (4.4,-1.13) coordinate (a_2);
				\draw[-] (4,-0.17) coordinate (a_1) -- (4.4,-1.13) coordinate (a_2);
				
				\begin{pgfonlayer}{background}
					\path[fill=gray!40!white,rounded corners, draw=black!10] (-1,-1.8) -- (5.3,-1.8) -- (5.3,.4) -- (-1,.4) -- (-1,-1.8);
				\end{pgfonlayer}
			\end{tikzpicture}
		}
	\end{center}
	\vspace{-1em}
	\caption{\footnotesize{Factor graph of six users ($J=6$) and four resource nodes
			($K = 4$) where  three users  connect with one resource node.}}
	\label{fig::factor_graph1}
\end{figure}
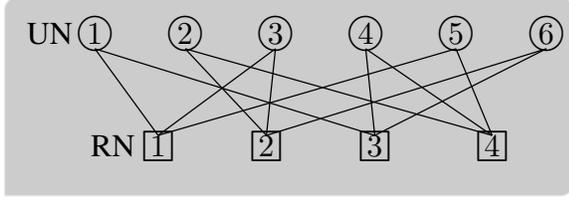

The codewords in SCMA are sparse, and only a few components of a codeword are non-zero.
This feature can be specified in terms of a $K \times J$ matrix $F$ called the factor graph matrix.  For example, consider $F$ shown  in (\ref{F_mat}) for an SCMA system with $J=6, K=4$ and $\lambda=150 \%$.
\begin{equation}
	\label{F_mat}
	F=
	\begin{bmatrix}
		1 & 0 & 1 & 0 & 1 & 0\\
		0 & 1 & 1 & 0 & 0 & 1\\
		1 & 0 & 0 & 1 & 0 & 1\\
		0 & 1 & 0 & 1 & 1 & 0\\
	\end{bmatrix}
\end{equation}
The  $1$s present in the $j$th column  specify the  locations of non-zero components of the codewords for the $j$th user. Since, the codewords are sparse, $F$ is also sparse.
The matrix $F$ can also be graphically represented by a factor graph as shown in Fig.~\ref{fig::factor_graph1}.   Corresponding to each column  and each row of $F$, there is a user node and a resource node respectively. And, against a `1' in $F$, there is an edge between the corresponding  user node and resource node.  

\subsubsection{Detection Using MPA} \label{MPA}

The MUD  for SCMA is a crucial task.
The factor graph of the SCMA system is sparse.  Therefore, MUD can be carried out by applying   MPA over the sparse graph.
The MPA and  its variants have been extensively used in many applications like decoding of low-density parity-check (LDPC), detection in multiple-input multiple-output (MIMO) antenna systems etc.

In the following we describe the version of  MPA which is used in MUD for SCMA system.
The underlying principle of the  MPA is  the maximum {\textit{a posteriori}} rule. It is numerically exhaustive to perform the MAP detection for all the users in a block at one shot. Instead, user-wise MAP detection is carried out.  The user-specific rule is given by
\begin{equation}
	\label{MAP}
	{\hat{\bf{x}}}_j=\arg \max_{{\bf{x}}_{jm} \in {\cal{X}}_j} V_{j}\left({\bf{x}}_{jm}\right)
\end{equation}
where ${\cal{X}}_j$ is the  codebook dedicated for the $j^{\text{th}}$ user,  ${\bf{x}}_{jm}$  is the $m^{\text{th}}$ codeword of ${\cal{X}}_j$ and $V_{j}\left({\bf{x}}_{jm}\right)$ is the {a posteriori} probability of the $j^{\text{th}}$ user's codeword being ${\bf{x}}_{jm}$ given the received signal ${\bf{y}}=[y_1,y_2, \cdots, y_K]^T$ with $m=1,2,\cdots, M$.

\begin{figure}[htb!]
	\centering
	\scalebox{0.75}{\includegraphics[]{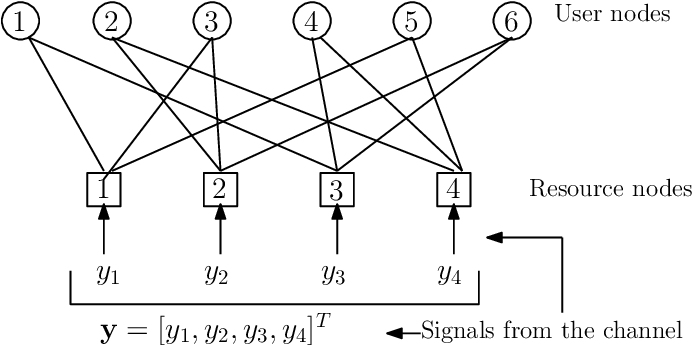}}
	\caption {MPA-based detection over factor graph }	 	
	\label{fig:det}
\end{figure}

The detection in (\ref{MAP}) is carried out iteratively by applying MPA. The detection process can be conveniently visualized graphically in terms of the factor graph as shown in Fig~\ref{fig:det}.  At the outset, the signals from the channel enter the resource nodes as shown in Fig.~\ref{fig:det}.  Note that unlike the LDPC codes where the user nodes (variable nodes) receive the channel values, here, the resource nodes (check nodes) receive the same.
Then the user nodes and the resource nodes exchange messages iteratively to extract the data of different users.
The steps for MPA-based MUD are presented \cite{sharma}	  below.

Let $U_{k\rightarrow j}^{(l)}$ and $V_{j\rightarrow k}^{(l)}$  be the messages sent by the $k$th resource node to the $j$th user node and by the   $j$th user node to the $k$th resource node, respectively,  during the $l$th iteration.  Here, the messages are vectors of length $M$. The $m$th  components   $U_{k\rightarrow j}^{(l)}\left({\bf{x}}_{jm}\right)$ and $V_{j\rightarrow k}^{(l)}\left({\bf{x}}_{jm}\right)$ correspond to the $m$th codeword ${\bf{x}}_{jm}$ present in the codebook ${\cal{X}}_j$.        The steps of the MPA are outlined below.
\begin{enumerate}
	\item {\textit{\underline{Initialization}}}  : At the beginning, the messages are equally-likely. Therefore, set $V_{j\rightarrow k}^{(0)}\left({\bf{x}}_{jm}\right)=1/M $. Note that in the case of the LDPC codes, the initialization is done according to the received channel values. Here, since the user nodes don't receive the signals from the channel, the messages (probability values) from the user nodes are set uniformly over the all possible symbols.
	\item {\textit{\underline{Resource node update}}} : The resource nodes receive the output signals $\bf{y}$ from the channel. The messages from the resource nodes are updated as follows:
	
	\begin{equation}
		\label{sumproduct}
		\begin{split}
			U_{k \rightarrow j}^{(l)}({\bf{x}}_{jm})& = \sum_{{\bf{c}} \in {\cal{C}}_{kj} } \frac{1}{\pi N_0} \exp [-\frac{1}{N_0}| y_k -\sqrt{P_j}h_{kj}x_{jmk} \\
			& - \sum_{j' \in {\cal{M}}_k^j} \sqrt{P_{j'}}h_{ki}c_{j'k} |^2] \prod_{j' \in {\cal{M}}_k^j} V_{j'\rightarrow k}^{(l-1)}\left({\bf{c}}_{j'}\right)
		\end{split}
	\end{equation}
	where,
	\begin{itemize}
		\item ${\cal{M}}_k$ is the set of the user nodes connected to the $k$th resource node and ${\cal{M}}_k^j={\cal{M}}_k \setminus \{j\}$.
		\item Suppose ${\cal{M}}_k^j=\left\{j_1, j_2, \ldots, j_{d_f-1}\right\}$. Then,   ${\cal{C}}_{kj}$ is the Cartesian product of the codebooks as defined below:
		$$ {\cal{C}}_{kj}= {\cal{X}}_{j_1}  \times {\cal{X}}_{j_2}  \times \cdots \times {\cal{X}}_{j_{d_f-1}} . $$
		\item ${\bf{c}} =\left({\bf{c}}_{j_1}, {\bf{c}}_{j_2}, \ldots, {\bf{c}}_{j_{d_f-1}}\right)$ is a member of ${\cal{C}}_{kj}$   and ${\bf{c}}_{j'}=\left(c_{j'1}, c_{j'2}, \ldots, c_{j'K}\right)$.
		\item ${\bf{x}}_{jm}=\left(x_{jm1}, x_{jm2}, \ldots, x_{jmK}\right)$.
	\end{itemize}
	The step in (\ref{sumproduct})  is known as the sum-product rule. The AWGN channel model is assumed.
	\item {\textit{\underline{User node update}}} :
	The messages from the user nodes are updated as follows:
	\begin{equation}
		\label{V_update}
		V_{j\rightarrow k}^{(l)}\left({\bf{x}}_{jm}\right) =\prod_{k' \in {\cal{N}}_j^k} U_{k' \rightarrow j}^{(l)}({\bf{x}}_{jm}) 
	\end{equation}
	where, ${\cal{N}}_j$ is the set of the  resource nodes connected to the $j$th user node and ${\cal{N}}_j^k={\cal{N}}_j \setminus \{k\}$.
	\item {\textit{\underline{Stopping rule}}} :
	If the messages get converged or the maximum number of iterations $\kappa$ is exhausted, then stop and proceed for decision making in Step (5). Otherwise, set $l=l+1$ and go to Step (2).
	\item {\textit{\underline{Decision}}} :
	Compute
	\begin{equation}
		\label{posterior_update}
		V_{j}\left({\bf{x}}_{jm}\right)=\prod_{k \in {\cal{N}}_j} U_{k \rightarrow j}^{(l)}({\bf{x}}_{jm}).
	\end{equation}
	The estimates of the transmitted codewords are found as follows:
	$$ {\hat{\bf{x}}}_j=\arg \max_{{\bf{x}}_{jm} \in {\cal{X}}_j} V_{j}\left({\bf{x}}_{jm}\right). $$
	
	The probability-domain MPA is usually numerically unstable. Therefore, the messages must be normalized in each iteration.
\end{enumerate}

\subsection{Example}
Now we give a complete example of the SCMA system. We explain the procedure for SCMA encoding and detection. Suppose, the code books   shown in Table~\ref{table:codebook1} 
are considered. 
\begin{table}[ht]
	\centering
	\caption{Codebooks of  $6$ users}
	\label{table:codebook1}
	\scalebox{0.6}{$
		\begin{gathered}
			{\cal{X}}_1=\left\{\begin{bmatrix}
				-1.2078 \\
				0\\
				-0.1339-0.3623i\\
				0
			\end{bmatrix}
			\begin{bmatrix}
				-0.4022 \\
				0\\
				0.7247 + 1.2078i\\
				0
			\end{bmatrix}
			\begin{bmatrix}
				0.4022 \\
				0\\
				-0.7247 - 1.2078i\\
				0
			\end{bmatrix}
			\begin{bmatrix}
				1.2078 \\
				0\\
				0.1339+0.3623i\\
				0
			\end{bmatrix}  \right\} \\
			{\cal{X}}_2=\left\{\begin{bmatrix}
				0\\
				-0.1339 - 0.3623i  \\
				0\\
				-1.2078
			\end{bmatrix}
			\begin{bmatrix}
				0\\
				0.7247 + 1.2078i \\
				0\\
				-0.4022
			\end{bmatrix}
			\begin{bmatrix}
				0\\
				-0.7247 - 1.2078i\\
				0\\
				0.4022
			\end{bmatrix}
			\begin{bmatrix}
				0\\
				0.1339 + 0.3623i\\
				0\\
				1.2078
			\end{bmatrix}   \right\} \\
			{\cal{X}}_3=\left\{\begin{bmatrix}
				-0.7247 - 1.2078i\\
				0.3623 - 0.3623i\\
				0\\
				0
			\end{bmatrix}
			\begin{bmatrix}
				-0.1339 - 0.3623i\\
				-0.7247 + 1.2078i \\
				0\\
				0
			\end{bmatrix}
			\begin{bmatrix}
				0.1339 + 0.3623i\\
				0.7247 - 1.2078i\\
				0\\
				0
			\end{bmatrix}
			\begin{bmatrix}
				0.7247 + 1.2078i\\
				-0.3623 + 0.3623i\\
				0\\
				0
			\end{bmatrix}   \right\} \\
			{\cal{X}}_4=\left\{\begin{bmatrix}
				0\\
				0\\
				-1.2078\\
				0.3623 - 0.3623i
			\end{bmatrix}
			\begin{bmatrix}
				0\\
				0\\
				-0.4022\\
				-0.7247 + 1.2078i
			\end{bmatrix}
			\begin{bmatrix}
				0\\
				0\\
				0.4022\\
				0.7247 - 1.2078i
			\end{bmatrix}
			\begin{bmatrix}
				0\\
				0\\
				1.2078\\
				-0.3623 + 0.3623i
			\end{bmatrix}   \right\} \\
			{\cal{X}}_5=\left\{\begin{bmatrix}
				0.3623 - 0.3623i \\
				0\\
				0\\
				-0.7247 - 1.2078i
			\end{bmatrix}
			\begin{bmatrix}
				-0.7247 + 1.2078i\\
				0\\
				0\\
				-0.1339 - 0.3623i 
			\end{bmatrix}
			\begin{bmatrix}
				0.7247 - 1.2078i\\
				0\\
				0\\
				0.1339 + 0.3623i
			\end{bmatrix}
			\begin{bmatrix}
				-0.3623 + 0.3623i \\
				0\\
				0\\
				0.7247 + 1.2078i
			\end{bmatrix}   \right\} \\
			{\cal{X}}_6=\left\{\begin{bmatrix}
				0\\
				-1.2078\\
				0.3623 - 0.3623i\\
				0
			\end{bmatrix}
			\begin{bmatrix}
				0\\
				-0.4022\\
				-0.7247 + 1.2078i \\
				0
			\end{bmatrix}
			\begin{bmatrix}
				0\\
				0.4022 \\
				0.7247 - 1.2078i \\
				0
			\end{bmatrix}
			\begin{bmatrix}
				0\\
				1.2078\\
				-0.3623 + 0.3623i \\
				0
			\end{bmatrix}   \right\}
		\end{gathered}
		$}
\end{table}

Consider the following data symbols for the 6 users:  $\left(2, 2, 1, 1, 3, 4\right)$. Then, the set of the codewords chosen for transmission  can be presented as shown below:
$$
\scalebox{0.5}{$\begin{pmatrix}
	-0.4022  &   0&   -0.7247 - 1.2078i &   0 &    0.7247 - 1.2078i & 0 \\
	0 &   0.7247 + 1.2078i &  0.3623 - 0.3623i &    0  &   0  & 1.2078 \\
	0.7247 + 1.2078i &   0  &   0 &-1.2078  &   0  & -0.3623 + 0.3623i\\
	0 &  -0.4022 &    0 &    0.3623 - 0.3623i &   0.1339 + 0.3623i & 0
\end{pmatrix}$}
$$

In the above, the $i$th column of the matrix represents the codeword transmitted by the $i$th user. Note that here the number of resources is $K=4$  and therefore the length of each codeword is also 4.  The sum of the superimposed  codewords is given by:
$$
\small
{\bf{r}}=
\begin{pmatrix}
	-0.4022 - 2.4156i \\
	2.2948 + 0.8454i\\
	-0.8454 + 1.5701i\\
	0.0941 + 0.0000i
\end{pmatrix}.
$$

We consider a simple complex AWGN channel model. Suppose, the AWGN noise samples at $\frac{E_b}{N_0}=3$dB  are given by:
$$
\small
{\bf{n}}=
\begin{pmatrix}
	0.0018 - 1.2008i \\
	1.2143 + 0.4227i  \\
	0.3323 + 0.4232i  \\
	0.1488 - 0.6993i
\end{pmatrix}
$$

Then the 4-dimensional received signal ${\bf{y=r+n}}$ becomes 
$$
\small
{\bf{y}}=
\begin{pmatrix}
	-0.4004 - 3.6164i \\
	3.5091 + 1.2681i\\
	-0.5132 + 1.9933i\\
	0.2428 - 0.6993i\\
\end{pmatrix}
$$
Given this value of ${\bf{y}}$, the individual user's data symbol is to be  estimated now.  For that we consider the probability-domain MPA described in Section~\ref{MPA}.    In the MPA-based SCMA detection process, the messages are vectors of length $M$  (in this case $M=4$).  There is no exchange of messages between a user node and a resource node if they are not connected through an edge. The vector messages for a particular direction in the entire factor graph can be stored in a matrix of size $KM\times J$.    First, the  messages from the user nodes  are initialized equally-likely as described in Section~\ref{MPA}.  These initialized messages are stored in the matrix $\bf{V}$ as shown below.   
\begin{equation}
	\label{V_initial}
	\small
	{\bf{V}}=
	\begin{pmatrix}
		0.25 &        0  &   0.25    &     0  &   0.25     &    0 \\
		0.25  &        0  &    0.25   &       0   &  0.25    &       0  \\
		0.25   &      0  &   0.25  &        0  &   0.25     &       0  \\
		0.25   &       0  &   0.25   &         0   &    0.25  &         0 \\ \hline
		0  &  0.25  &   0.25    &      0   &       0   &  0.25  \\
		0  &  0.25  &  0.25    &       0   &        0   &  0.25  \\
		0  &   0.25   &   0.25  &       0     &    0  &  0.25 \\
		0  &  0.25  &  0.25    &     0  &        0   &   0.25 \\  \hline
		0.25   &       0   &       0  &   0.25   &       0 &    0.25\\
		0.25   &      0   &      0  &  0.25 &        0 &    0.25\\
		0.25  &       0    &      0  &   0.25  &         0  &   0.25  \\
		0.25  &       0    &     0  &  0.25   &       0  &  0.25  \\ \hline
		0  &  0.25  &        0 &   0.25 &   0.25   &      0  \\
		0  &   0.25  &        0  &   0.25  &  0.25   &         0\\
		0  &  0.25 &         0  &  0.25  &   0.25  &        0  \\
		0  &  0.25    &       0   &  0.25  &    0.25 &         0 
	\end{pmatrix}
\end{equation}
Observe from (\ref{V_initial})  that $\bf{V}$ contains $K=4$ blocks of rows of size $M=4$. The number of columns is $J=6$. 
After the initialization step,  the messages from the resource nodes are updated according to  (\ref{sumproduct}). These messages are stored in a matrix $\bf{U}$ as shown below:
\begin{equation}
	\label{U_msg}
	\small
	{\bf{U}}=
	\begin{pmatrix}
		
		0.1764 &        0 &    0.9962   &      0  &  0.0042  &       0 \\
		0.6415  &       0  &  0.0038   &      0  &  0.0000  &        0 \\
		0.1784  &        0 &   0.0000   &        0  &    0.9958  &         0 \\
		0.0038  &       0  &  0.0000    &      0  &   0.0000  &        0 \\ \hline
		0   &  0.0000   & 0.7860    &     0   &       0  &  0.0000 \\
		0   &  0.9986  &   0.0000    &     0  &        0  &   0.0000 \\
		0   &  0.0000  &   0.2042  &        0   &       0   &  0.0079  \\
		0   &  0.0014  &   0.0098  &        0 &        0  &  0.9921  \\  \hline
		0.0328 &         0  &       0   &  0.2660  &        0 &    0.0168 \\
		0.6324   &       0  &        0  &   0.4763  &         0   &  0.6641  \\
		0.0002  &       0    &      0   &  0.2362  &        0  &    0.0000  \\
		0.3347   &      0    &     0   &   0.0215   &        0   &   0.3191  \\  \hline
		0  &  0.1512  &       0   &  0.3688   &  0.2511  &        0  \\
		0  &    0.3453  &        0  &  0.0387  &   0.3669  &        0 \\
		0   & 0.2673   &       0   & 0.3406  &   0.3087  &        0 \\
		0   &  0.2362  &        0  &  0.2519 &   0.0733   &      0
	\end{pmatrix}
\end{equation}

In order to demonstrate the resource node update, we consider the calculation of $U_{1\rightarrow 1}(1)$ during the first iteration.  Note that the 1st resource node is connected to user nodes 1, 3 and 5.  Since we are calculating the 1st component of $U_{1\rightarrow 1}$,  the data symbol $v_1$ for the 1st user node is fixed at 1. There are $M^{d_f-1}=4^2=16$ different combinations for $v_3$ and $v_5$.  Table~\ref{table:res_update}  shows the detailed calculation for each of the combinations.  

\begin{table}[ht]
	\caption{Computation of $U_{1\rightarrow 1}(1)$:  A total of 16 terms/combinations in the sum} 
	\centering 
	\begin{tabular}{c c c c} 
		\hline\hline 
		$v_1$ & $v_3$ & $v_5$ & Term \\ [0.5ex] 
		\hline 
		1 & 1& 1 & 6.1e-7 \\ 
		1 & 1 & 2 & 7.1e-18 \\
		1 & 1 & 3 & 6.1e-4 \\
		1 & 1 & 4 & 6.9e-12 \\
		1 & 2& 1 & 1.2e-9 \\ 
		1 & 2 & 2 & 8.8e-22 \\
		\bf{1} &\bf{2} & \bf{3} & \bf{8.5e-6} \\
		1 & 2 & 4 & 6.2e-15 \\
		1 & 3& 1 & 1.5e-13\\ 
		1 & 3 & 2 & 3.9e-27 \\
		1 & 3 & 3 & 8.8e-9 \\
		1 & 3 & 4 & 2.2e-19 \\
		1 & 4& 1 & 1.9e-19 \\ 
		1 & 4 & 2 & 3.2e-34 \\
		1 & 4 & 3 & 8.1e-14 \\
		1 & 4 & 4 & 1.3e-25 \\
		\hline 
		$U_{1\rightarrow 1} (1)$  (sum) &  & & 6.2e-4
	\end{tabular}
	\label{table:res_update} 
\end{table}

As an example consider the computation of the combination corresponding to $v_1=1, v_3=2, v_5=3$.  Here $x_{111}=-1.2078$, $c_{32}=-0.1339 - 0.3623i$ and $c_{53}=0.7247 - 1.2078i$. At $\frac{E_b}{N_0}=3$dB, we have $N_0=0.5012$.  With these values, the term is computed as:
$$
\small
\begin{aligned}
	{\text{term}}& =\frac{1}{\pi N_0}\exp \left[-\frac{1}{N_0}\left|y_1-x_{111}-c_{32}  -c_{53}\right|^2\right]\times V_{3\rightarrow 1}(2)\times V_{5\rightarrow 1}(3)\\
	&= \frac{1}{0.5 \pi}\exp \left[-\frac{1}{0.5}\left|-0.4 - 3.6i +1.2+0.1 + 0.36i -0.7 + 1.2i\right|^2\right] (0.25)^2\\
	&=8.5e-6
\end{aligned}
$$
Finally, the   terms corresponding to all the combinations are summed up to produce $U_{1\rightarrow 1}(1)=6.2e-4$.   The values of $U_{1\rightarrow 1}(2)$, $U_{1\rightarrow 1}(3)$ and $U_{1\rightarrow 1}(4)$ are  $0.0022, 6.2471e-4$ and $1.33e-5$ respectively. These components are normalized so that we have $\small {\bf{U}}_{1\rightarrow 1}=[0.1764, 0.6415, 
0.1784, 0.0038]^T$.

After the computation of all the messages from the resource nodes, the messages from the user nodes are updated according to (\ref{V_update}).  These messages are shown below:

$$
\small
{\bf{V}}=
\begin{pmatrix}
	0.0328 &        0  &   0.7860    &     0  &   0.2511     &    0 \\
	0.6324  &        0  &    0.0000   &       0   &  0.3669    &       0  \\
	0.0002   &      0  &   0.2042  &        0  &   0.3087     &       0  \\
	0.3347   &       0  &   0.0098   &         0   &    0.0733  &         0 \\
	0  &  0.1512  &   0.9962    &      0   &       0   &  0.0168  \\
	0  &  0.3453  &  0.0038    &       0   &        0   &  0.6641  \\
	0  &   0.2673   &   0   &       0     &    0  &  0 \\
	0  &  0.2362  &  0.0000    &     0  &        0   &   0.3191  \\
	0.1764   &       0   &       0  &   0.3688   &       0 &    0\\
	0.6415   &      0   &      0  &  0.0387 &        0 &    0\\
	0.1784  &       0    &      0  &   0.3406  &         0  &   0.0079  \\
	0.0038  &       0    &     0  &  0.2519   &       0  &  0.9921  \\
	0  &  0  &        0 &   0.2660 &   0.0042   &      0  \\
	0  &   0.9986  &        0  &   0.4763  &  0    &         0\\
	0  &  0 &         0  &  0.2362  &   0.9958  &        0  \\
	0  &  0.0014    &       0   &  0.0215  &    0 &         0 
\end{pmatrix}
$$
Then the {\textit{a posteriori}} probability vectors are updated as per (\ref{posterior_update}). These values are shown below: 

$$
\small
{{\bf{V}}_{\text{posterior}}}=
\begin{pmatrix}
	0.0058 &    0.0000  &  0.7830 &     0.0981 &    0.0010  &   0.0000\\
	0.4056 &   0.3448  &   0.0000  &    0.0184 &    0.0000  &   0.0000 \\
	0.0000 &   0.0000  &  0.0000  &  0.0804  &  0.3074  &  0.0000  \\
	0.0013 &   0.0003  &   0.0000 &   0.0054 &   0.0000  &   0.3166
\end{pmatrix}
$$
Then by identifying the indices of the maximum component in the columns, the  user's symbols are estimated as $2,2,1,1,3,4$. 
Note that the normalization must be applied only to $\bf{U}$  and $\bf{V}$ as these will be used further in the next round of iterations. Usually there is no concrete stopping rule as in the case of LDPC codes where a zero syndrome vector terminates the iterations.  Instead, here, the algorithm is stopped when there is no significant changes in ${{\bf{V}}_{\text{posterior}}}$  or the maximum number of iterations are exhausted. 
\subsection{Numerical Results and Discussions}
In this section, the numerical results are shown for an SCMA system.

In Fig. \ref{fig1}, average SER performance of SCMA system is shown.  We consider SCMA system performance using various codebooks for comparison in
Fig. \ref{fig1}.  Further,  in Fig. \ref{fig1} the SCMA
system's SER performance for nine users ($J = 9$) and six
resource elements ($K = 6$) over AWGN channel  is considered.  Therefore, SCMA system's performance depends on the design of codebooks of each user as observed in Fig. \ref{fig1}. The gain using the codebook in \cite{sharma}  is around 1 dB as observed in  Fig. \ref{fig1}.

\pgfplotsset{every semilogy axis/.append style={
		line width=0.7 pt, tick style={line width=0.7pt}}, width=8cm,height=7cm, 
	legend style={font=\scriptsize},
	legend pos= north east}
\begin{figure}[!htb]
	\centering
	\begin{tikzpicture}
		\begin{semilogyaxis}[xmin=0, xmax=10, ymin=.9e-4, ymax=1,
			xlabel={SNR (dB)},
			ylabel={ SER},
			grid=both,
			grid style={dotted},
			legend cell align=left,
			legend entries={{Sharma \cite{sharma}}, {Zhang \cite{zhang}},{Yu \cite{yu}},{Nikopour, \cite{nikopor}} },
			cycle list name=black white
			]
			\addplot [cyan, mark=otimes, mark size=3 pt]table [x={x1}, y={y4}] {6by9.txt};	
			\addplot [green, mark=diamond, mark size=3 pt]table [x={x1}, y={y3}] {6by9.txt};
			\addplot [black, mark=o, mark size=3 pt] table [x={x1}, y={y2}] {6by9.txt};
			\addplot [red, mark=square, mark size=2 pt]table [x={x1}, y={y1}] {6by9.txt};
			
		\end{semilogyaxis}
	\end{tikzpicture}
	\caption{{BER performance of the SCMA system using various codebooks for
			$J = 9$ and $K = 6$ (overloading factor $\lambda=150\%$ ) in AWGN channel.}}
	\label{fig1}
\end{figure}
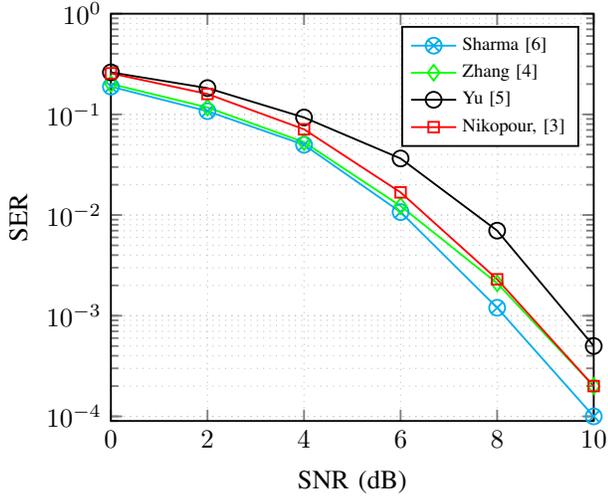


Further, $6 \times 4$,   $8 \times 4$ and  $9 \times 6$ SCMA system's SER performance is shown in Fig. \ref{awgn}  over AWGN channel. $8 \times 4$ SCMA system has worst performance due to high overloading factor as compared to  $6 \times 4$ and  $9 \times 6$ SCMA systems.
Further, $9 \times 6$
SCMA system has better performance than the $6 \times 4$ due to
less symbol collisions among the users. Furthermore, same results are observed over Rayleigh fading channel in Fig. \ref{rayleigh}.

\pgfplotsset{every semilogy axis/.append style={
		line width=0.7 pt, tick style={line width=0.7pt}}, width=8cm,height=7cm, 
	legend style={font=\scriptsize},
	legend pos= north east}
\begin{figure}[!htb]
	\centering
	\begin{tikzpicture}
		\begin{semilogyaxis}[xmin=0, xmax=12, ymin=.9e-4, ymax=1,
			xlabel={SNR (dB)},
			ylabel={ SER},
			grid=both,
			grid style={dotted},
			legend cell align=left,
			legend entries={{$6 \times4$ SCMA}, {$8 \times4$ SCMA},{$9 \times6$} SCMA},
			cycle list name=black white
			]
			\addplot [red, mark=square, mark size=3 pt]table [x={x1}, y={4by6}] {awgn.txt};	
			\addplot [blue, mark=diamond, mark size=3 pt]table [x={x1}, y={4by8}] {awgn.txt};
			\addplot [black, mark=o, mark size=3 pt] table [x={x1}, y={6by9}] {awgn.txt};
			
		\end{semilogyaxis}
	\end{tikzpicture}
	\caption{{BER performance of the SCMA system using various
			$J $ and $K$  in AWGN channel.}}
	\label{awgn}
\end{figure}
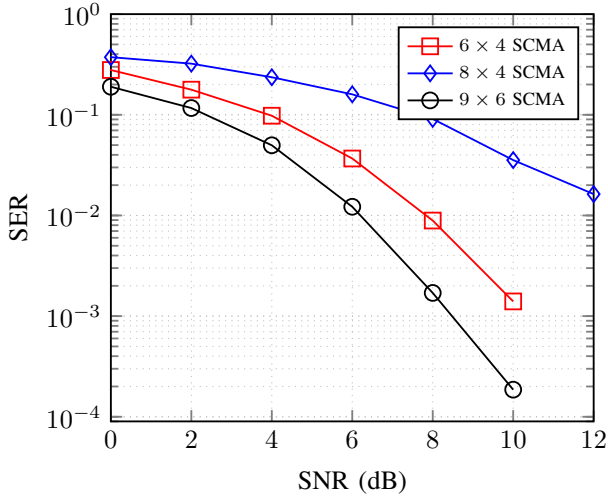

\pgfplotsset{every semilogy axis/.append style={
		line width=0.7 pt, tick style={line width=0.7pt}}, width=8cm,height=7cm, 
	legend style={font=\scriptsize},
	legend pos= north east}
\begin{figure}[!htb]
	\centering
	\begin{tikzpicture}
		\begin{semilogyaxis}[xmin=0, xmax=18, ymin=1e-4, ymax=1,
			xlabel={SNR (dB)},
			ylabel={ SER},
			grid=both,
			grid style={dotted},
			legend cell align=left,
			legend entries={{$6 \times4$ SCMA}, {$8 \times4$ SCMA},{$9 \times6$} SCMA},
			cycle list name=black white
			]
			\addplot [red, mark=square, mark size=3 pt]table [x={x1}, y={4by6}] {rayleigh.txt};	
			\addplot [blue, mark=diamond, mark size=3 pt]table [x={x1}, y={4by8}] {rayleigh.txt};
			\addplot [black, mark=o, mark size=3 pt] table [x={x1}, y={6by9}] {rayleigh.txt};
			
		\end{semilogyaxis}
	\end{tikzpicture}
	\caption{{BER performance of the SCMA system using various
			$J $ and $K$  in Rayleigh channel.}}
	\label{rayleigh}
\end{figure}
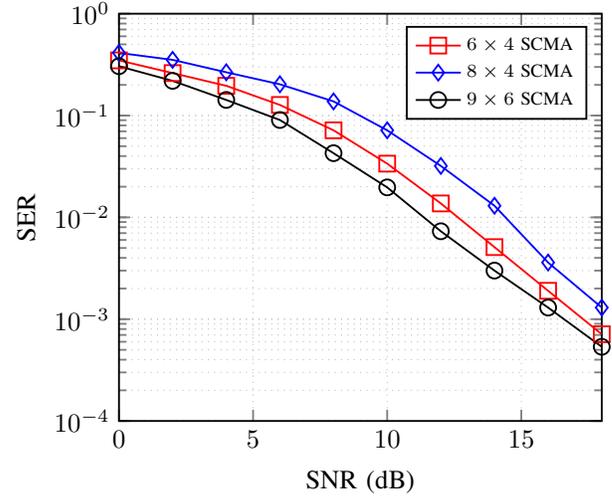

\section{Hybrid Multiple Access-based System Design}
In this section, we discuss a hybrid multiple access (HMA) approach to further enhance the spectral efficiency of next-generation wireless network.
In real scenario, users' distribution around the base station   is uniform. Further, based on users' channel gain,  users are grouped  into near users (NU) and far users (FU) in a cell.  FU have higher pathless than NU in system.  Furthermore, the difference can
also be possible between the number of FU (more) and the NU (less) in a system due to a higher area of peripheral than the near area
of a base station. Therefore, a conventional PD-NOMA approach is not optimal for  pairing all the FU and NU in a cell. Since, some FU are left after NU and FU pairing, as shown in Figure \ref{hybrid}. Further, the SCMA based method to connect all the users to a base station is not optimal since it requires multiple SCMA group, where each SCMA group uses orthogonal resources. Therefore, a HMA scheme is considered to connect all the users in a system using minimum number of resources.

In HMAS, users are divided into multiple groups based on their channel gain difference. For example, in a downlink scenario, users are partitioned into two groups, say Group-1 and Group-2, as shown in Fig. \ref{hybrid}.
Group-1 and Group-2 are near and far to a base station and is referred to as
near group (or strong group) and far group (or weak group), respectively.
Let $J_1$ and $J_2$ users  in Group-1 and Group-2, respectively.

The strong and the weak users in HMA-based system are denoted by $\{s_1, s_2, \ldots, s_{J_1}\}$  and  $\{w_1, w_2, \ldots, w_{J_2}\}$, respectively.
Let $\left\{{\bf{h}}_i^{s}\right\}_{i=1}^{J_1}$ and  $\left\{{\bf{h}}_j^{w}\right\}_{j=1}^{J_2}$ denote the channel impulse response vectors experienced by the strong and the weak users, respectively.
Therefore,  $J_1 \times K$ and $J_2 \times K$  SCMA systems for {\textit{Group 1}} based on a codebook ${\mathcal{C}}_1$ and  {\textit{Group 2}}, based on a codebook ${\mathcal{C}}_2$, are  considered, respectively. 
Let ${\bf{x}}^s_i \in \mathcal{C}_1$ and ${\bf{x}}^w_j \in \mathcal{C}_2$ denote the codewords for the $i$th strong user and the  $j$th weak user respectively.
The superimposed the codewords of all   users at the transmitter  is written as:
$$
{\bf{x}}=  \sum_{i=1}^{J_1} \sqrt{P^s_i}{\bf{x}}^s_i  + \sum_{j=1}^{J_2}\sqrt{P^w_j}{\bf{x}}^w_j
$$
where, $ P^s_i$ and $ P^w_j$ denote the powers assigned to the $i$th strong and $j$th weak user respectively.
The received signal vectors at the  strong and the weak users are given by
\begin{equation}
	\begin{aligned}
		{\bf{y}}_i^s& =\text{diag}\left({\bf{h}}_i^s\right){\bf{x}}  +{\bf{n}}_i^s,   \;\;\;\; i=1,2, \ldots J_1 \text{  and  }\\
		{\bf{y}}_j^w& =\text{diag}\left({\bf{h}}_j^w\right){\bf{x}}  +{\bf{n}}_j^w,  \;\;\;\; j=1,2, \ldots J_2
	\end{aligned}
	\label{comb_eq}
\end{equation}
respectively.

\begin{figure}[!htbp]
	\begin{center}
		\tikzstyle{block} = [draw, fill=blue!25, rectangle, text centered, minimum height=3em, text width=4em]
		\tikzstyle{sum} = [draw, fill=blue!20, circle, node distance = 1cm]
		\tikzstyle{line} = [draw,-stealth,thick]
		\begin{tikzpicture}[scale=0.8, every node/.style={transform shape}]
			
			\node[block,text width=4.3em, xshift=0, yshift= 0em](qam){{MPA Detector (MPAD2)}};
			
			\node  [left of=qam,node distance=2cm](1){$\textbf{y}^s_i$};
			\node  [above of=1,node distance=1.2cm, xshift=2em](sig){};
			
			\node[block,right of = qam,node distance=3cm, xshift=0em, yshift=0em,text width=4.8em](recond){{Re-encoder}};	
			
			\node [sum,right of = recond, node distance=2cm](plus1){\tiny{${\sum}$}};
			\node [left of = plus1, xshift=1.6em, yshift=-.6em](plus11){\large{$-$}};
			\node [right of = plus1, xshift=-2.1em, yshift=1.2em](plus111){\large{$+$}};
			
			\node[block,right of = plus1,node distance=2cm, xshift=0.5em, yshift=0em,text width=4.8em](encod2){{MPA Detector (MPAD1)}};
			\node[right of = encod2,node distance=1.5cm, xshift=0em, yshift=0em](pp1){};
			\node[below of = pp1,node distance=1.4cm, xshift=0em, yshift=0em](pp2){};
			\node [below of = qam, xshift=4em, yshift=-1.3em](weak1){{Far users' data }};
			\node [below of = recond, xshift=12.65em, yshift=-1.3em](weak){{Near users' data }};
			\node [above of = recond, node distance=1.14cm,xshift=2em, yshift=1.3em](sic){SIC};
			\node [left  of = encod2, xshift=-0.5em, yshift=1em](interference){{${\bf{y}}^s_{i, \text{SIC}}$ }};
			\path[line](encod2)--(pp1);
			\path[line](sig)-|(plus1);
			\path[line](1)--(qam);	
			\path[line](qam)--(recond);	
			\path[line](recond)--(plus1);
			\path[line](plus1)--(encod2);
			\draw[line] (1.5, 0) coordinate (a_1) -- (1.5, -1.2) coordinate (a_2);
			\draw[line] (8.55, 0) coordinate (a_1) -- (8.55, -1.25) coordinate (a_2);
			\draw[line] (-1.21, 0) coordinate (a_1) -- (-1.21, 1.24) coordinate (a_2);
			
			\draw[dotted, thick] (1.7,-.7) rectangle (5.6,1.4);

			\begin{pgfonlayer}{background}
				\path[fill=gray!40!white,rounded corners, draw=black!50] (-2.2,-2.1) -- (8.7,-2.1) -- (8.7,1.8) -- (-2.2,1.8) -- (-2.2,-2.1);
			\end{pgfonlayer}
			
		\end{tikzpicture}
	\end{center}
	\caption{Detection at the location of the strong/near users \cite{deka}.}
	\label{receiver1}
\end{figure}
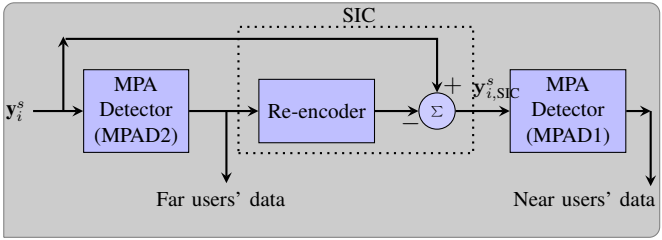

\begin{figure*}[htb!]
	\centering
	\includegraphics[width=\linewidth]{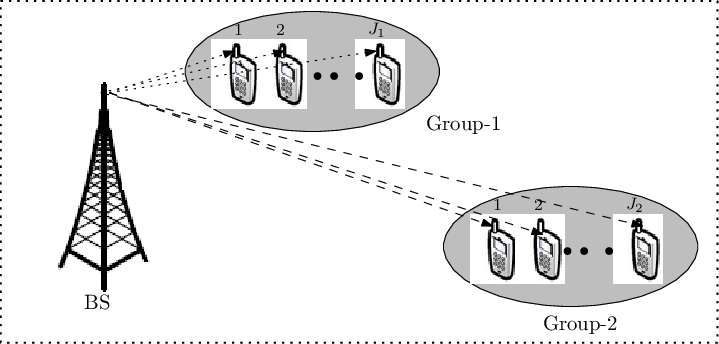}
	\caption{Hybrid  Multiple access  system. }	 	
	\label{hybrid}
\end{figure*}

MPA detector for the users in \textit{Group 1} operates over the factor graph corresponding to the codebook ${\cal{C}}_1$. Such an MPA detector is denoted by MPAD1. Similarly, the MPA detector for \textit{Group~2} users is referred to as MPAD2. First,  the detection of the strong users is explained. The detection of the $i$th strong user $s_i$ is illustrated in Fig.~\ref{receiver1}. The SIC of the far users is performed before extracting the data of $s_i$. First, with ${\bf{y}}^s_i$ as input, MPAD2 produces the estimated codewords $\left\{\hat{{\bf{x}}}^{w_i}_j\right\}_{j=1}^{J_2}$. Here, the superscript `$i$' signifies that the detection process is carried out by the $i$th receiver  at the location of $s_i$. These estimates are  used only for detecting the data of $s_i$. The weak users' signal is recreated and the interference-canceled received signal is obtained as
$$
{\bf{y}}^s_{i, \text{SIC}} ={\bf{y}}_i^s-  \sum_{j=1}^{J_2}\text{diag}\left({\bf{h}}_j^w\right)\sqrt{P^w_j}\hat{{\bf{x}}}^{w_i}_j.$$
Using ${\bf{y}}^s_{i, \text{SIC}}$ as the input, the MPA detector MPAD1 estimates the data of $\left\{s_1, \ldots, s_{J_1}\right\}$. From these, the estimate for only $s_i$ is considered while ignoring the rest.

The detection of the weak users is carried out similarly  as in the  conventional downlink SCMA system.  Observe that, in the proposed system, the overloading factor is  increased from $\lambda= \frac{J_{c}}{K}$ to  $\lambda_{\mathrm {proposed}}= \frac{J_1+J_2}{K}$ where, $J_c$ is the number of users in a conventional SCMA system.

HMAS-based  system's average SER  performance is shown in Fig.
\ref{hybrid1} with conventional SCMA and PD-NOMA-based systems' performance.
HMA and PD-NOMA-based system involve detection based on
SIC principle, which suffers from error propagation.
Therefore,
these systems are outperformed by the purely SCMA-based
systems,which
involve powerful MPA detection. HMA-based system has better performance than conventional PD-NOMA or SCMA system with a higher overloading factor as observed in Fig. \ref{hybrid1}.

\pgfplotsset{every semilogy axis/.append style={
		line width=0.7 pt, tick style={line width=0.7pt}}, width=8cm,height=7cm, 
	legend style={font=\scriptsize},
	legend pos= north east}
\begin{figure}[!htb]
	\centering
	\begin{tikzpicture}
		\begin{semilogyaxis}[xmin=0, xmax=30, ymin=1e-4, ymax=1,
			xlabel={SNR (dB)},
			ylabel={ SER},
			grid=both,
			grid style={dotted},
			legend cell align=left,
			legend entries={ {Hybrid, $J=12, K=4 (\lambda=300\%)$}, {PD-NOMA, $\lambda=200\%$},{SCMA, $J=8, K=4 (\lambda=200\%)$}}
			]
			\addplot [green!50!black, mark=o, mark size=2 pt]table [x={x1}, y={S1 }] {fading_new.txt};
			\addplot [red, mark=square, mark size=2 pt] table [x={x1}, y={S2}] {fading_new.txt};
			\addplot [blue, mark=triangle, mark size=2 pt]table [x={x1}, y={S3}] {fading_new.txt};
			
		\end{semilogyaxis}
	\end{tikzpicture}
	\caption{{SER over Rayleigh  channel}}
	\label{hybrid1}
\end{figure}
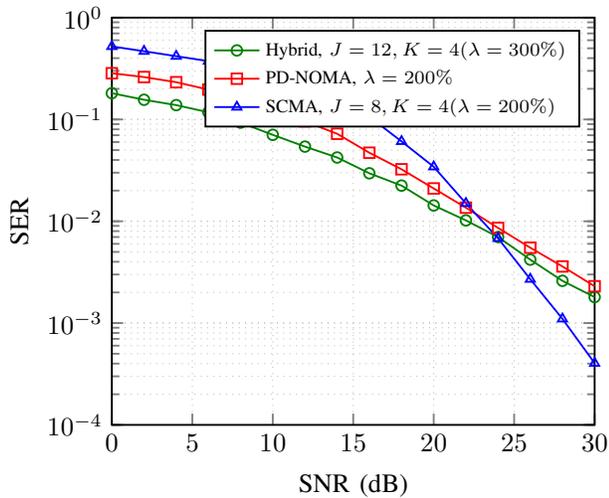

\subsection{Future Research Direction}
SCMA and HMA schemes can be used to improve the spectral efficiency of 5G and beyond networks. However, some design and implementation issues have to be solved  for practical implementation  of SCMA and HMA schemes such as follows:

\begin{itemize}
	\item SCMA or HMA-based system's performance depends on the codebooks assigned to the each user. Therefore, an efficient and universal method of codebooks design is essential. However, currently, codebooks are designed using suboptimal approaches and also these design depends on the overloading factor in the system.
	\item In general,  MPA-based detection algorithms are used in SCMA or HMA. However, MPA has higher complexity for higher modulation order size and/or resource degree of each node in factor graph. Therefore, some simple algorithms are essential for practical implementation of SCMA-based system at higher data rate.
	
	\item An optimal power allocation scheme is also required to split the total power among the groups in HMA-based system. Since the total sum rate of HMA system can be enhanced using optimal power allocation.
	\item In practice,  the perfect channel information is not available at receiver. Therefore, effect of imperfect channel should be analyzed in SCMA or HMA-based system for detection and estimation.
	\item Further, massive multi-input and multi-output  (MIMO) will be an integral part of 5G and beyond  wireless network. Therefore, SCMA and HMA methods should be analyzed with massive MIMO system.
	
\end{itemize}

\end{document}